\documentclass[twocolumn,final,twoside,journal]{IEEEtran}

\usepackage{tabu}
\usepackage{xcolor}

\usepackage{amsmath,amsthm,graphicx,cite}
\usepackage{srcltx}
\usepackage{epsfig,amsfonts}
\usepackage{graphicx,cite,amssymb,amsmath}
\usepackage{color}

\usepackage{amsthm}
\usepackage[nolist]{acronym}
\usepackage{psfrag}
\usepackage{perso}
\usepackage{pgfplots}
 \pgfplotsset{compat=newest}
    \pgfplotsset{plot coordinates/math parser=false}
    \pgfplotsset{
    label style={anchor=near ticklabel},
    xlabel style={yshift=0.0em},
    ylabel style={yshift=-0.3em},
    tick label style={font=\footnotesize },
    label style={font=\footnotesize},
    legend style={font=\footnotesize},
    title style={font=\fontsize{7}}}
\definecolor{iso}{rgb}{0.7,0.7,0.7}
\usepackage{blindtext}
\usepackage{relsize}
\usepackage{mathtools}
\mathtoolsset{showonlyrefs}

\usepackage{placeins}

\ifCLASSOPTIONcompsoc
\usepackage[caption=false,font=normalsize,labelfont=sf,textfont=sf]{subfig}
\else
\usepackage[caption=false,font=footnotesize]{subfig}
\fi

\usepgflibrary{arrows}
\usetikzlibrary{patterns}
\usepgflibrary{decorations.pathmorphing}

\newcommand{\rippleset}{\mathscr{R}}
\newcommand{\cloudset}{\mathscr{C}}

\newcommand{\Ripple}{\mathtt{R}}
\newcommand{\Cloud}{\mathtt{C}}
\renewcommand{\c}{\mathtt{c}}
\renewcommand{\r}{\mathtt{r}}

\newcommand{\Ripp}{\mathsf{R}}

\newcommand{\ripple}[1]{ \msr{R}_{#1}}
\newcommand{\cloud}[1]{\msr{C}_{#1}}

\newcommand{\ru}{\r_u}
\newcommand{\Ru}{\Ripple_u}
\newcommand{\Cu}{\Cloud_u}
\newcommand{\cu}{\c_u}

\renewcommand{\S}[1]{\mathtt{S}_{#1}}
\newcommand{\s}[1]{\pmb{\mathtt{s}}_{#1}}
\newcommand{\w}{\pmb{\mathtt{w}}}

\ifthenelse{\isundefined{\C}}{
    \newcommand{\C}[1]{\mathtt{C}_{#1}}
    }{
    \renewcommand{\C}[1]{\mathtt{C}_{#1}}
    }

\ifthenelse{\isundefined{\N}}{
    \newcommand{\N}[1]{\mathsf{N}_{#1}}
    }{
    \renewcommand{\N}[1]{\mathsf{N}_{#1}}
    }

\newcommand{\Erv}{\mathtt{A}}
\newcommand{\erv}{\mathtt{a}}
\renewcommand{\b}{\mathtt{b}}
\newcommand{\B}{\mathtt{B}}
\newcommand{\n}{\mathtt{n}}

\renewcommand{\nu}{\n_u}

\newcommand{\bu}{\b_u}
\newcommand{\Bu}{\B_u}

\newtheorem{mydef}{Definition}

\newtheorem{theorem}{Theorem}

\newcommand{\paccess}{p}
\newcommand{\pu}{q_u}

\newcommand{\slot}{y} 

\newcommand{\per}{{\mathsf{P}}_{\mathsf{e}}}
\newcommand{\throughput}{\mathsf{T}}
\newcommand{\nuser}{n}
\newcommand{\nslot}{m}

\newcommand{\cs}{k}

\newcommand{\myru}{l}
\newcommand{\myerv}{s}

\newcommand{\rdeg}{\text{red}}
\renewcommand{\deg}{\text{deg}}

\usepackage{multirow}

\begin{document}

\begin{acronym}
\acro{RA}{Random Access}
\acro{LT}{Luby Transform}
\acro{BP}{belief propagation}
\acro{i.i.d.}{independent and identically distributed}
\acro{PER}{Packet Error Rate}
\acro{SIC}{Successive Interference Cancellation}
\end{acronym}

\title{Finite-Length Analysis of Frameless ALOHA \\ with Multi-User Detection}

\author{
    \IEEEauthorblockN{Francisco L\'azaro\IEEEauthorrefmark{1}, \v Cedomir Stefanovi\'c\IEEEauthorrefmark{2},~\IEEEmembership{Member,~IEEE}\\
}

\thanks{F. L\'azaro is with the Institute of Communications and Navigation of DLR (German Aerospace Center), Wessling, Germany. Email: Francisco.LazaroBlasco@dlr.de.}
\thanks{\v Cedomir Stefanovi\' c is with the Department of Electronic Systems, Aalborg University,
Aalborg, Denmark. Email: cs@es.aau.dk). His work was supported by the Danish Council for Independent Research, grant no. DFF-4005-00281.}
\thanks{\copyright 2016 IEEE. Personal use of this material is permitted. Permission
from IEEE must be obtained for all other uses, in any current or future media, including
reprinting /republishing this material for advertising or promotional purposes, creating new
collective works, for resale or redistribution to servers or lists, or reuse of any copyrighted
component of this work in other works}
}


\maketitle



\thispagestyle{empty} \pagestyle{empty}

\begin{abstract}
In this paper we present a finite-length analysis of frameless ALOHA for a $\cs$ multi-user detection scenario, i.e., assuming the receiver can resolve collisions of size $\cs$ or smaller.
The analysis is obtained via a dynamical programming approach, and employed to optimize the scheme's performance.
We also assess the optimized performance as function of $k$.
Finally, we verify the presented results through Monte Carlo simulations.
\end{abstract}

\vspace{-0.6cm}
\section{Introduction}\label{sec:Intro}

Slotted ALOHA (SA) \cite{R1975} is a widely used random access protocol, where users randomly and independently select slots in which they transmit their packets to a common access point (AP).
Frequently, SA is analyzed using a collision channel model, where a collision of two or more packets is considered destructive (i.e., all involved packets are lost), while slots that contain a single packet (singleton slots) are always successfully decoded. In this setting, the maximum expected throughput of SA is $1/e$.

\begin{figure}[t]\centering
	\includegraphics[width=0.48\columnwidth]{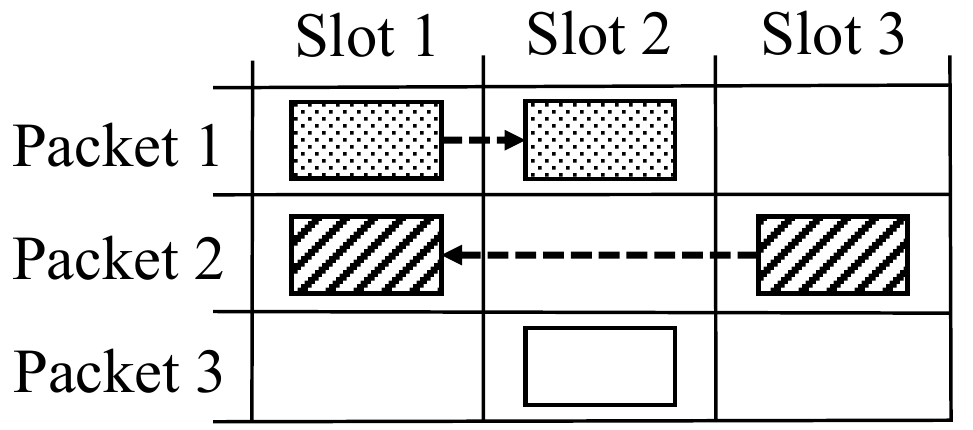}
	\caption{Example of SIC-enabled slotted ALOHA: Packet 2 is received in singleton slot 3, recovered and its replica cancelled from slot 1. Slot 1 now becomes singleton, packet 1 is recovered and its replica cancelled from slot 2. Slot 2 becomes singleton and packet 3 is recovered from it.}
	\label{fig_1}
    \vspace{-0.4cm}
\end{figure}

The introduction of \ac{SIC} in SA framework significantly changed the perspective on the capabilities of random access protocols \cite{CGH2007}.
Namely, assume that a user sends replicas of the same packet in multiple slots, embedding in each replica pointers to the slots where the other replicas are sent.
A packet occurring in a singleton slot is successfully received, enabling the identification of the slots containing the other replicas and their removal via \ac{SIC}, see Fig.~\ref{fig_1}.
This may turn some of the collided slots into singletons, propelling the recovery of new packets and the removal of their replicas.
This process is analogous to the iterative belief-propagation erasure-decoding, promoting the use of theory and tools of codes-on-graphs to design and analyze SA schemes \cite{L2011}.
In this way, the asymptotic throughput for the collision channel model can be pushed to the ultimate limit of 1 packet per slot \cite{PSLP2014}.
These insights inspired a strand of works that applied various concepts from codes-on-graphs to SIC-enabled SA; we refer the reader to \cite{PSLP2014} for an overview.
In this paper we focus on frameless ALOHA \cite{SPV2012,SP2013}, which exploits ideas originating from the rateless coding framework \cite{luby02:LT}.
Frameless ALOHA is characterized by (i) a contention period that consists of a number of slots that is not defined a priori, but terminated when the number of resolved users\footnote{Under user resolution we assume recovery/decoding of user packet.} and/or instantaneous throughput reach certain thresholds and (ii) a slot access probability with which a user decides on a slot basis whether to transmit a packet or not.

In this paper we consider a $\cs$ multi-user detection (MUD) setting in which the AP is able to decode collisions of size up to  $\cs$ at the receiver \cite{SG2006}, which can be understood as a generalization of the collision channel model.
Building up on an approach devised for rateless codes \cite{Karp2004,lazaro:Allerton2015}, we advance the theoretical treatment of SIC-enabled SA schemes by providing an exact finite-length analysis of frameless ALOHA with $\cs$-MUD.
We use the analysis to optimize the performance by maximizing the expected throughput, and show that the maximum expected throughput\footnote{We define the throughput as the number of resolved user normalized by the number of slots \emph{and} by $\cs$, see Section~\ref{sec:finite_L_MUD}.} does not depend on $\cs$.
The results are verified via Monte-Carlo simulations.

The rest of the paper is organized as follows.
This section is concluded with a brief overview of the related work.
Section~\ref{sec:sysmodel} describes the system model.
Section~\ref{sec:finite_L_MUD} presents the proposed finite-length analysis, while the performance optimization is assessed in Section~\ref{sec:opt}.
Finally, Section~\ref{sec:Conclusions} concludes the paper.
\vspace{-0.9cm}
\subsection*{Related Work}

The asymptotic performance optimization of frameless ALOHA was done in \cite{SPV2012}, while the joint assessment of the optimal slot access probability and the contention termination criteria in finite-length scenarios via simulation in \cite{SP2013}.
An approximate finite-length analysis of the performance of irregular repetition SA \cite{L2011} in the error floor region was done in \cite{ivanov:floor}.
Examples of works analyzing and optimizing the performance of classical SA (i.e., without SIC) with $\cs$-MUD can be found in \cite{NTS2008,GSG2009,GS2011}.
Further, the asymptotic analysis of irregular repetition SA \cite{L2011} in $\cs$-MUD scenario was presented in \cite{GS2013}.
Finally, the finite-length analysis of slotted ALOHA for the standard collision channel model was reported in \cite{lazaro:SCC2017}.
This work extends the analysis in \cite{lazaro:SCC2017} to the $\cs$-collision channel, i.e., for the case in which $\cs$-MUD is employed at the receiver, and presents  numerical results highlighting how the throughput depends on the MUD capabilities of the receiver.


\section{System Model}\label{sec:sysmodel}

\begin{figure}[t]
        \centering
        \subfloat {\includegraphics[width=0.37\columnwidth]{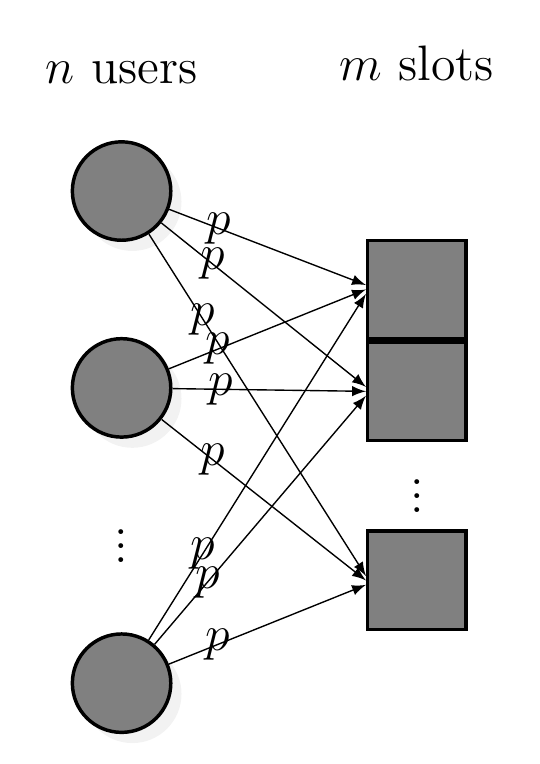}
        \label{fig_2a} }
        \vline
        \subfloat { \includegraphics[width=0.49\columnwidth]{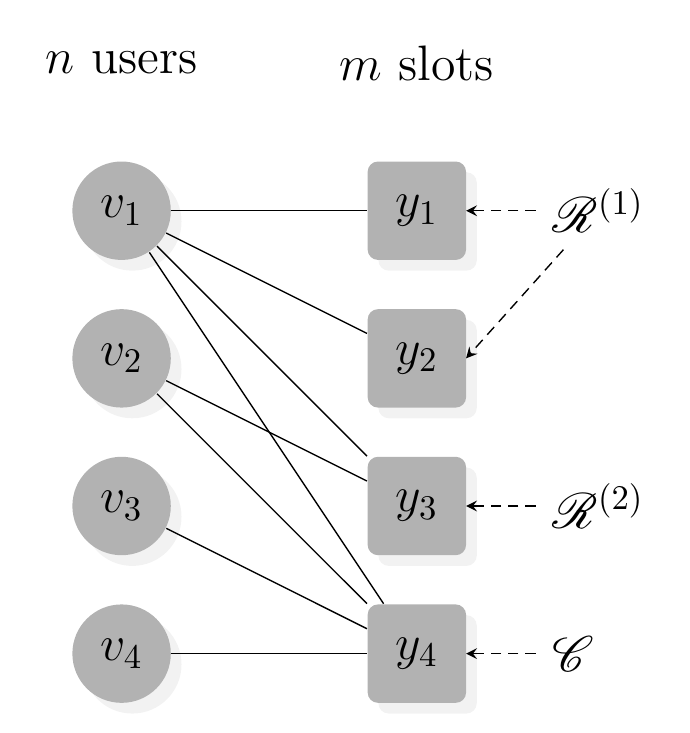}
        \label{fig_2b} }
        \caption{(a) Contention model. (b) Example of cloud and ripples, $k=2$.}
        \label{fig_2}
        \vspace{-0.4cm}
\end{figure}

We consider a single instance of batch arrival of $\nuser$ users, contending for the access to the AP.
The contention period is assumed to have a duration of $\nslot$ slots ($\nslot$ is not defined a-priori but determined on the fly), and users are assumed to be slot and contention period synchronous, all arriving prior to the start of the contention period.
A user contends by transmitting replicas of the same packet; for each slot of the contention period the user decides with slot access probability $\paccess$ whether to transmit a replica, independently of any other slot and of any other user, as shown in Fig~\ref{fig_2a}.
For the sake of simplicity, we assume that $\paccess$ is uniform over users and slots and equal to
\begin{align}
\paccess = \frac{\beta}{\nuser}
\end{align}
where $\beta$ is a suitably chosen constant.
Denoting with $\Omega_i$ the probability of a slot having degree $i$, it is easy to verify that
\[
\Omega_i = \binom{\nuser}{i} \paccess^i (1-\paccess)^{\nuser-i}, \; i = 0,\ldots,n.
\]


%


The decoding process at the AP is described using a bipartite graph. 
The users are denoted by $v_i$, $ i =1, \ldots, \nuser$, and the slots by $\slot_j$, $ j = 1, \ldots, \nslot$.
The notation $\deg(\slot)$ is used to refer to the (original) degree of a slot, i.e., the number of users that transmitted in the slot.
We also introduce the term reduced degree to refer to the number of unresolved packets that are still present in the slot during the decoding, and denote it by $\rdeg(\slot)$, where $\rdeg(\slot) \leq \deg(\slot)$.

The $\cs$-MUD is modeled such that slots containing up to $\cs$ transmissions are decoded with probability $1$, while slots containing more than $\cs$ transmissions are undecodable with probability 1, c.f. \cite{NTS2008,GSG2009,GS2011}; for $k=1$, this model reduces to the standard collision channel model.
For ease of analysis, we assume that the decoder decodes and removes through SIC exactly $1$ transmission per iteration.\footnote{The assumption has no impact on the derived performance, but only models the operation of the $k$-MUD receiver in a way that is consistent with the framework in \cite{Karp2004,lazaro:Allerton2015,lazaro:SCC2017}.}
Thus, when the decoder is applied to a slot of degree $h \leq \cs$, it performs $h$ iterations, each time reducing the slot degree by one. Concretely, if there are several slots with degrees up to $\cs$, the slot with minimum degree $h$ among them is chosen; if there are several slots with degree $h$, one of them is chosen at random.
Once a slot is chosen, one of the $h$ colliding users is selected at random, decoded and removed from the slot, and from all the other slots where the replicas occurred.

We introduce the following definitions:
\begin{mydef}[$h$-th Ripple] We define the $h$-th ripple as the set of slots of reduced degree $h$ and we denote it by $\rippleset^{(h)}$.
\end{mydef}
\noindent The cardinality of the $h$-th ripple is denoted by $\r^{(h)}$ and its associated random variable as $\Ripp^{(h)}$. 
\begin{mydef}[Cloud] We define the cloud as the set of slots with reduced degree $d > \cs$ and we denote it by $\cloudset$.
\end{mydef}
\noindent The cardinality of the cloud is denoted by $\c$ and the corresponding random variable as $\Cloud$.

Fig.~\ref{fig_2b} shows an example of bipartite graph for $\nuser=4$ users and $\nslot=4$ slots for $\cs=2$.
Observe that slots $\slot_1$ and $\slot_2$ belong to the first ripple $\rippleset^{(1)}$, slot $\slot_3$ belongs to the second ripple $\rippleset^{(2)}$, and $\slot_4$ belongs to the cloud $\cloudset$.


Finally, we add a temporal dimension to the cloud and ripples through the subscript $u$ that corresponds to the number of unresolved users.
Initially, all $\nuser$ users are unresolved, hence $u=\nuser$.
At each iteration, if the ripple is not empty, exactly 1 user gets resolved and the subscript decreases by 1.
Decoding ends successfully (all users are decoded) if $u=0$, or
unsuccessfully if at any of the $\nuser$ decoding steps there is no slot whose (reduced) degree is less then or equal to $\cs$.


\section{Finite-Length Analysis}
\label{sec:finite_L_MUD}

Following the approach in \cite{Karp2004,lazaro:Allerton2015,lazaro:SCC2017}, the iterative decoding of frameless ALOHA in the $\cs$-MUD scenario is represented as a finite state machine with state
\[
\S{u}:=( \Cu, \Ru^{(\cs)}, \Ru^{(\cs-1)}, \cdots, \Ru^{(1)} )
\]
i.e., the state comprises the cardinalities of the cloud and the $\cs$-th to first ripples at the decoding step in which $u$ users are unresolved.
The following theorem establishes a recursion that can be used to determine the decoder state distribution.
\begin{theorem}\label{theorem:rec_mud}
Given that the decoder is at state ${\S{u}=(\cu,\ru^{(\cs)},\ru^{(\cs-1)},\cdots, \ru^{(1)} )}$, when $u$ users are unresolved and $\sum_{i=1}^{\cs} \ru^{(i)}>0$ (i.e., at least one ripple being non-empty), the probability of the decoder being at state ${\Pr \{\S{u-1}= \s{u-1}\}}$ when $u-1$ users are unresolved is given by
\begin{align}
\Pr\{ \S{u-1} &= (  \s{u} + \w )| \S{u} =  \s{u} \} =
 \binom{\cu}{\bu} {\pu}^{\bu} (1-\pu)^{\cu-\bu} \\[-4ex]
& \times \mathlarger{\prod}_{h=1}^{h=\cs} {\binom{\myru^{(h)}}{\myerv^{(h)}} \left(\frac{h}{u}\right)^{\myerv^{(h)}} \left( 1- \frac{h}{u} \right)^{\myru^{(h)}-\myerv^{(h)}} }
\label{eq:prob_transition}
\end{align}
with
\[\s{u} =(\cu, \ru^{(\cs)}, \ru^{(\cs-1)}, \cdots, \ru^{(1)} ) \]
\[{\w =(-\bu, \bu-\erv_u^{(\cs)}, \erv_u^{(\cs)} -\erv_u^{(\cs-1)}, \cdots,\erv_u^{(2)} -\erv_u^{(1)}  ) }\]
and
\vspace{-2ex}
\begin{align}
\pu =  \frac{ \mathlarger {\sum}\limits_{d=k+1}^{\nuser}   \Omega_d \, \frac{d}{\nuser}  \binom{d-1}{\cs}  \frac{\binom{u-1}{\cs}}{\binom{\nuser-1}{\cs}}  \frac{\binom{\nuser-u}{d-\cs-1}}{\binom{\nuser-\cs-1}{d-\cs-1}} }
{  1 -  \mathlarger{\sum}\limits_{h=0}^{\cs} \,\,
\mathlarger{\sum}\limits_{d=h}^{\nuser}    \Omega_d \, \frac{\binom{u}{h} \binom{\nuser-u}{d-h}}{\binom{\nuser}{d}} }
\label{eq:pu_theorem_mud}
\end{align}
\begin{align}
  \myru^{(h)}  = \begin{cases}
                   \ru^{(h)}-1, & \mbox{if } \ru^{(h)}>0  \text{ and }  \sum_{i=1}^{h-1} \ru^{(i)}=0\\
                    \ru^{(h)}, & \mbox{otherwise}.
                 \end{cases} \\
  \myerv^{(h)}  = \begin{cases}
                   \erv_u^{(h)}-1, & \mbox{if } \ru^{(h)}>0  \text{ and }  \sum_{i=1}^{h-1} \ru^{(i)}=0\\
                    \erv_u^{(h)}, & \mbox{otherwise}
                 \end{cases}
\end{align}
for
\vspace{-2ex}
\begin{align}
\erv_u^{(h)}-\erv_u^{(h+1)} & \leq \ru^{(h)}, \, h=1,2,\cdots, \cs-1 \\
\erv_u^{(\cs)} - \bu  &\leq \ru^{(\cs)} \text{ and } 0 \leq \bu \leq \cu .
\end{align}
\end{theorem}

\begin{IEEEproof}
The proof consists of analyzing the variation of the cloud and the ripple cardinalities in the transition from $u$ to $u-1$ unresolved users.
Since we assume that $\sum_{i=1}^{\cs} \ru^{(i)}>0$, exactly one user is resolved in the transition and all edges connected to the resolved user are erased from the decoding graph.
As a consequence, some slots might leave $\cloudset_u$ and enter $\rippleset_{u-1}^{(\cs)}$, leave $\rippleset_{u}^{(\cs)}$ and enter $\rippleset_{u-1}^{(\cs-1)}$ etc.

We focus first of the number of slots leaving $\cloud{u}$ and entering $\ripple{u-1}^{(\cs)}$ in the transition, denoted by $\bu$ and the associated random variable $\Bu$.
Due to the nature of frameless ALOHA, it can be assumed that a slot chooses its neighbor users uniformly at random and without replacement.
Thus, random variable $\Bu$ is binomially distributed with parameters $\c_u$ and $\pu$, being $\pu$ the probability of a generic slot $\slot$ leaving $\cloud{u}$ to enter $\ripple{u-1}^{(\cs)}$,
\begin{equation}
\pu = \Pr \{ \slot \in \ripple{u-1}^{(\cs)} | \slot \in \cloud{u} \}= \frac { \Pr \{ \slot \in \ripple{u-1}^{(\cs)}\, , \, \slot \in \cloud{u} \} }  { \Pr \{ \slot \in \cloud{u} \}}.
\label{eq:pu_prob_mud}
\end{equation}
We evaluate the numerator in \eqref{eq:pu_prob_mud} conditioning on the degree of slot $\slot$, i.e., via $\Pr \{ \slot \in \ripple{u-1}^{(\cs)} ,\, \slot \in \cloud{u} | \deg(\slot)= d \}$.
This corresponds to the probability that exactly one of the $d$ edges of  slot $\slot$ is connected to the user being resolved at the transition, out of the remaining $d-1$ edges, exactly $\cs$ edges are connected to the $u-1$ unresolved users after the transition, and the remaining $d-\cs-1$ edges are connected to the $\nuser-u$ unresolved users before the transition.
This probability is
\begin{align}
\label{eq:z_and_l_d_mud}
\Pr \{ \slot \in \ripple{u-1}^{(\cs)} ,\, \slot \in \cloud{u} |  \deg(\slot) = d \} =  \\
\begin{cases}
\frac{d}{\nuser} \binom{d-1}{\cs}  \frac{\binom{u-1}{\cs}}{\binom{\nuser-1}{\cs}}  \frac{\binom{\nuser-u}{d-\cs-1}}{\binom{\nuser-\cs-1}{d-\cs-1}}, & d > k
 \\
0, & d  \leq k
\end{cases}
\end{align}
since for $d < \cs$, the slot cannot enter the $k$-th ripple.

We now turn to the denominator in \eqref{eq:pu_prob_mud}, i.e., the probability that a slot $\slot$ is in the cloud when $u$ users are unresolved.
This corresponds to the probability that the reduced degree of $\slot$ is neither equal nor smaller than $\cs$, which can be casted as
\vspace{-1ex}
\begin{align}
\Pr   \{ \slot & \in  \cloud{u}\} =   1 - \sum\limits_{h=0}^{\cs} \Pr \{ \rdeg_u(\slot) = h \}  \\[-1ex]
& = 1 - \sum\limits_{h=0}^{\cs}  \sum\limits_{d=0}^{n} \Omega_d  \Pr \{ \rdeg_u(\slot) = h | \deg(\slot)= d\}
\label{eq:z_mud}
\end{align}
where the term  $\Pr \{ \rdeg_u(\slot) = h | \deg(\slot)= d\}$ corresponds to the probability of a slot $\slot$ with (original) degree $d$, having exactly $h$ neighbors among the $u$ unresolved users and $d-h$ neighbors among the $k-u$ resolved users:
\vspace{-2ex}
\begin{equation}
\Pr \{  \rdeg_u(\slot) = h | \deg(\slot)= d\}  =
\mathlarger{\sum}\limits_{d=h}^{\nuser}    \Omega_d \, \frac{\binom{u}{h} \binom{\nuser-u}{d-h}}{\binom{\nuser}{d}}
\label{eq:red_deg_denom}
\end{equation}
Inserting \eqref{eq:red_deg_denom},  \eqref{eq:z_mud} and \eqref{eq:z_and_l_d_mud} in \eqref{eq:pu_prob_mud}, we obtain $\pu$ in \eqref{eq:pu_theorem_mud}, and thereby determine the variation of size of the cloud, i.e., random variable $\Bu$.

We focus next on the number of slots that leave the $\cs$-th to first ripple in the transition from $u$ to $u-1$.
We denote by $\erv_u^{(h)}$ the number of slots leaving the $h$-th ripple in the transition from $u$ to $u-1$ unresolved users, and refer to the associated random variable as $\Erv_u^{(h)}$.
We distinguish between two cases.
In the first case, the ripples $1$ to $h-1$ are empty, but the $h$-th ripple is not empty.
Thus, ${\r_{u}^{(h)}\geq1}$ and ${\sum_{i=1}^{h-1} \ru^{(i)}=0}$. One of the slots of the $h$-th ripple is selected at random and one of the involved users gets resolved.
In addition, the remaining ${\ru^{(h)}-1}$ slots in $\rippleset_{u}^{(h)}$  leave this ripple with probability $h/u$, which corresponds to the probability that they have the resolved user as neighbor.
Hence,
\begin{align}
\Pr\{\Erv_u^{(h)}=\erv_u^{(h)} & |\Ru^{(h)}=\ru^{(h)}\}=\\ &\binom{\ru^{(h)}-1}{\erv_u^{(h)}-1} \left(\frac{h}{u}\right)^{\erv_u^{(h)}-1} \mkern-3mu \left( 1- \frac{h}{u} \right)^{\ru^{(h)}-\erv_u^{(h)}}
\end{align}
for ${\r_{u-1}^{(h)}>1}$ and ${\sum_{i=1}^{h-1} \ru^{(i)}=0}$.
In the second case, the ripples $h-1$ to $1$ are not all empty, thus, ${\sum_{i=1}^{h-1} \ru^{(i)}>0}$, so one of the slots in $\rippleset_{u}^{(h-1)}$ to $\rippleset_{u}^{(1)}$ is used to resolve one user.
Therefore, any slot in $\rippleset_{u}^{(h)}$ in which the resolved user transmitted a replica will leave this ripple.
Since a slot in $\rippleset_{u}^{(h)}$ has reduced degree $h$ and the resolved user is selected at random from the $u$ unresolved users, we have
\begin{align}
\Pr\{\Erv_u^{(h)}=\erv_u^{(h)} & |\Ru^{(h)}=\ru^{(h)}\}=\\ &\binom{\ru^{(h)}}{\erv_u^{(h)}} \left(\frac{h}{u}\right)^{\erv_u^{(h)}} \mkern-3mu \left( 1- \frac{h}{u} \right)^{\ru^{(h)}-\erv_u^{(h)}}
\end{align}
for ${\sum_{i=1}^{h-1} \ru^{(i)}>0}$.

The proof is completed by observing that
\begin{align}
\c_{u-1} = \cu - \bu, \; \r_{u-1}^{(\cs)}= \ru^{(\cs)} -\erv_u^{(\cs)} + \bu \text{, and}  \\
\r_{u-1}^{(h)}= \ru^{(h)} -\erv_u^{(h)} + \erv_u^{(h+1)},  \, \, \forall \, h<\cs.
\end{align}

\end{IEEEproof}

The initial state of the decoder corresponds to a multinomial distribution over $\nslot$ slots and $k+2$ possible outcomes for each slot, corresponding to the slot being in the cloud, the $k$-th to first ripple or having degree 0. The probability of a slot belonging to the $h$-th ripple is given by $\Omega_h$, the probability of a slot having degree $0$ by $\Omega_0$ and the probability of a slot belonging to the $k$-th cloud by $1- \sum_{i=0}^{k} \Omega_i$.
Hence, we have
\begin{align}
& \Pr   \{\S{\nuser}=(\c_\nuser,\r_\nuser^{(\cs)},\r_\nuser^{(\cs-1)},\cdots, \r_\nuser^{(1)} ) \} =\\
&  \frac{\nslot!}{\c_\nuser! \, \r_\nuser^{(\cs)}! \, \r_\nuser^{(\cs-1)}! \cdots , \r_{\nuser}^{(1)}! \,  (\nslot-\c_{\nuser}-\r_{\nuser})!} \times \\
& \left( 1- \sum_{i=0}^{k} \Omega_i \right)^{\c_{\nuser}} \,   \Omega_\cs^{\r_{\nuser}^{(\cs)}} \,  \Omega_{\cs-1}^{\r_{\nuser}^{(\cs-1)}} \cdots \ ,\Omega_0^{\nslot-\c_\nuser -\sum_{i=1}^{\cs} \r_\nuser^{(i)} }
 \label{eq:init_state}
\end{align}
for all non-negative $\c_{\nuser},\r_{\nuser}^{(\cs)},\r_{\nuser}^{(\cs-1)},\cdots, \r_{\nuser}^{(1)}$, such that ${\c_{\nuser}+ \sum_{i=1}^{\cs}\r_{\nuser}^{(i)} \leq \nslot}$.

The decoder state probabilities are obtained by initializing the finite state machine according to \eqref{eq:init_state} and applying recursively Theorem~\ref{theorem:rec_mud}.
Once the decoder state probabilities are determined, it is possible to obtain the \ac{PER}, i.e., the probability that a user is not resolved when the decoding process ends, denoted by $\per$. Decoding ends at stage $u$ whenever $\sum_{i=1}^{\cs} \ru^{(i)}= 0$ (i.e., all ripples are empty), and this leaves exactly $u$ users unresolved. Thus,
\begin{align}
\per = \sum_{u=1}^{\nuser} \sum_{\cu} \frac{u}{\nuser} \Pr\{\S{u} =(\cu,0,0,\cdots,0) \}.
\end{align}
Finally, we define the expected throughput $\throughput$ as the number of resolved users normalized by $k$ and the number of slots\footnote{That is, we assume that $k$-MUD comes at the price of using $k$ times more time-frequency resources per slot, c.f. \cite{GS2013}.
A related result in \cite{GSP2014} shows that a combined use of $k$-out-of-$n$ signature-coding and lattice-coding provides $\cs$-MUD, with the required resources scaling, in essence, linearly with $\cs$.} 
\begin{align}
\throughput = \frac{\nuser \, (1-\per)}{k \, m}=\frac{1-\per}{ ( k \, \nslot )/\nuser}.
\label{eq:T}
\end{align}

\begin{figure}[t]
        \centering
        \subfloat {\includegraphics[height=3.7cm]{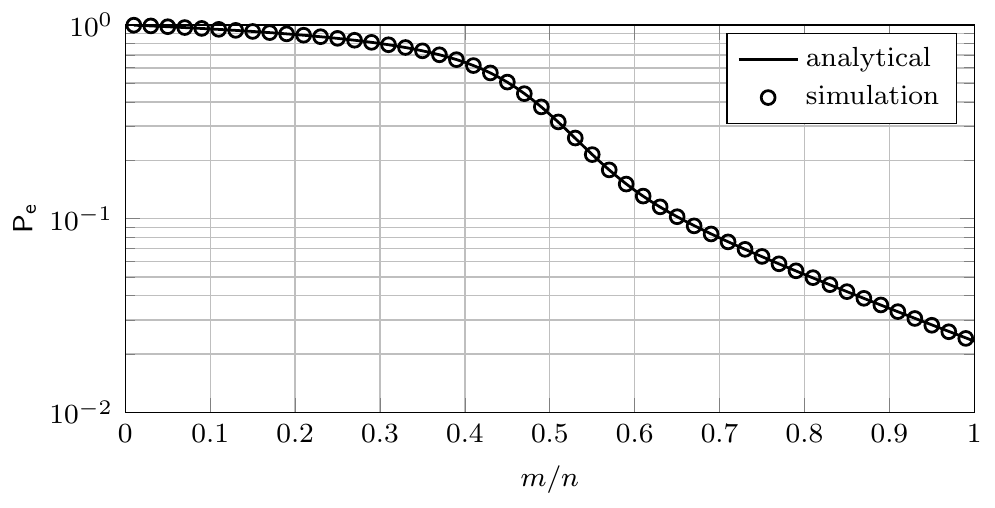}  }
        \vspace{-0.5cm} 
        \newline

        \hspace{-0.51cm}
        \subfloat { \includegraphics[height=3.68cm]{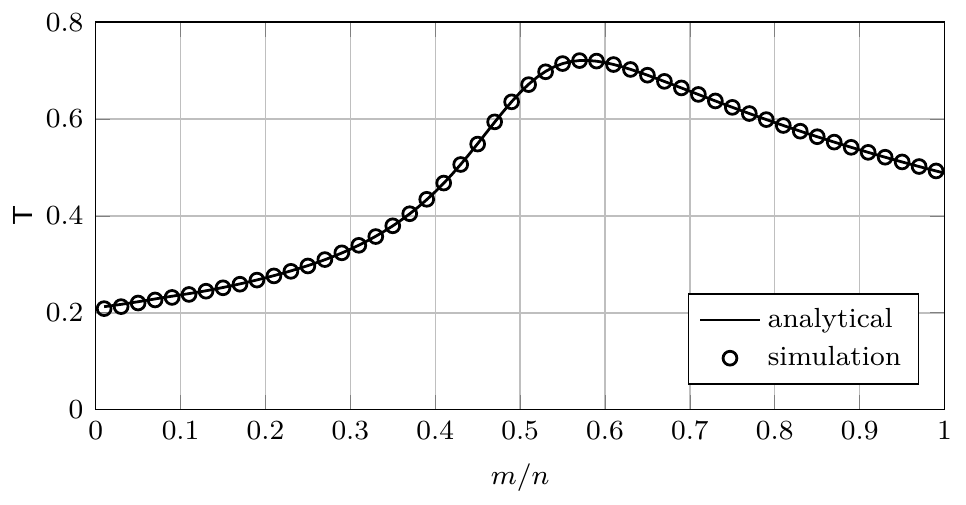} }
        \vspace{-0.4cm}
        \caption{Packet-error rate $\per$ and throughput $\throughput$ as a functions of $\nslot/\nuser$ for $\nuser=100$, $\cs=2$ and $\beta = 3.7$. } 
        \label{fig:example}
        \vspace{-0.4cm}
\end{figure}

In Fig.~\ref{fig:example} we show $\throughput$ and $\per$ as a function of $\nslot/\nuser$, for $n=100$, $\cs=2$ and $\beta=3.7$. The figure shows analytical results according to Theorem~\ref{theorem:rec_mud} and the outcome of Monte Carlo simulations. We see how the match is tight down to simulation error (10000 contentions periods were simulated).

\section{Optimization}\label{sec:opt}

In this section, we use the derived analysis to find the slot-access probability $\beta_{\text{opt}}$ that maximizes the peak expected throughput, $\throughput_{\max}$, for different values of $\nuser$ and $\cs$.

Table~\ref{tab:params_peak} lists the values of $\beta_{\text{opt}}$ obtained together with the peak expected throughput $\throughput_{\max}$ and the value of $\nslot / \nuser$ for which this maximum is achieved, for $\nuser=50, 100, 200$, and $\cs=1, 2, 3$.
Obviously,  $\beta_{\text{opt}}$ increases as $\nuser$ and/or $\cs$ increase.
Further, $\throughput_{\max}$ increases as $\nuser$ increases, but for fixed $\nuser$ it \emph{stays the same} as $\cs$ increases.
In other words, for the assumed simple scenario in which $\beta$ (i.e., $p$) is fixed on a slot basis,  our numerical results indicate that the throughput defined as in \eqref{eq:T} is not increased by increasing $\cs$. Thus investing in $\cs$-MUD does not pay off in terms of maximum expected throughput  $\throughput_{\max}$.
On the other hand, for fixed $\nuser$ as $\cs$ increases, the value of $\nslot / \nuser$ for which $\throughput_{\max}$ occurs decreases. Thus, investing in $\cs$-MUD may be of interest when one targets low latency and high throughput at the same time.

\begin{table}
\centering
\caption{Optimal parameters for frameless ALOHA with $\cs$-MUD}
\vspace{-0.1cm}
    \begin{tabular}{|c|c||c|c|c|}
    \hline
    $\nuser$				&  $\cs$	& $\beta_{\text{opt}} $ 	& $\throughput_{\max} $	& $ \nslot / \nuser \, (\throughput_{\max}) $ \\ \hline \hline
    \multirow{3}{*}{50}     & 1 	& 2.47				& 0.67				& 1.32	\\ \cline{2-5}
    						& 2		& 3.56				& 0.67				& 0.62	\\ \cline{2-5}
						    & 3		& 4.47				& 0.67				& 0.38	\\ \hline \hline
    \multirow{3}{*}{100}  	& 1		& 2.62				& 0.72				& 1.26 	\\ \cline{2-5}
    						& 2		& 3.81				& 0.72				& 0.58	\\ \cline{2-5}
						    & 3		& 4.86				& 0.72				& 0.36 	\\ \hline \hline
    \multirow{3}{*}{200} 	& 1		& 2.71				& 0.76				& 1.2 	\\ \cline{2-5}
    						& 2		& 4.04				& 0.76				& 0.56	\\ \cline{2-5}
						    & 3		& 5.22 				& 0.76				& 0.35	\\ \hline
    \end{tabular}
\label{tab:params_peak}
\vspace{-0.5cm}
\end{table}

\section{Conclusions and Discussion}\label{sec:Conclusions}
\vspace{-0.5ex}
In this paper we have presented an exact finite-length analysis of frameless ALOHA in the $\cs$-collision channel. The analysis is based on dynamical programming approach and is exact, both in the error floor region and in the waterfall region, as verified by means of Monte Carlo simulations.
The presented material can be extended to derive the asymptotic decoder behaviour by means of difference equations as done in \cite{Maatouk:2012}, which is the topic of our ongoing work.
\vspace{-2.0ex}

\bibliographystyle{IEEEtran}
\bibliography{IEEEabrv,references}

\end{document}